\documentclass[fleqn,twoside]{article}
\usepackage{espcrc2}

\usepackage{graphicx}
\usepackage[figuresright]{rotating}

\newcommand{\AmS}{{\protect\the\textfont2
  A\kern-.1667em\lower.5ex\hbox{M}\kern-.125emS}}

\newcommand{\Eq}[1]{Eq.(\ref{#1})}
\newcommand{\Fig}[1]{Fig.~\ref{#1}}      
\newcommand{\Ref}[1]{Ref.~\cite{#1}}

\newcommand{\Bs}{\hspace*{-1ex}}

\title{MEM Analysis of Glueball Correlators at $T>0$}

\author{
  Noriyoshi Ishii
  \address{Radiation Laboratory,
    RIKEN (The Institute of Physical and Chemical Research)\\
    2--1 Hirosawa, Wako, Saitama 351--0198, Japan}, and
  Hideo Suganuma
  \address{Graduate School of Science and Engineering, Tokyo Institute of Technology,\\
    Ohokayama 2--12--1, Meguro, Tokyo 152--8551, Japan}}

\begin{document}

\begin{abstract}
Maximum  Entropy Method (MEM)  is applied  to glueball  correlators at
finite  temperature constructed by  using 5,500  -- 9,900  gauge field
configurations  generated  in  SU(3)  quenched lattice  QCD  with  the
lattice  parameter  $\beta_{\rm  lat}  =  2N_c/g^2  =  6.25$  and  the
renormalized  anisotropy  $a_s/a_t  =  4$.  The  results  support  the
thermal width  broadening of the  $0^{++}$ glueball near  the critical
temperature.
\end{abstract}

\maketitle

It is  believed that, in  the neighborhood of the critical  temperature $T_c$, 
the QCD vacuum begins to change  its structure such  as the reduction  of the
string  tension, the  partial  restoration of spontaneous  chiral-symmetry  
breaking and  so on.   Since the  hadrons are  composites of
quarks and gluons, whose interactions  depend on the properties of the QCD
vacuum, some  of the  hadrons are expected  to change  their structure
drastically near $T_c$.
For instance, effective model studies suggest the pole-mass reductions
of  charmonium and  light  $q\bar{q}$ mesons  as  consequences of  the
reduction of  the string  tension and the  partial restoration  of 
spontaneous     breaking    of    chiral     symmetry,    
respectively~\cite{hashimoto,hatsuda,hatsuda2}.
In  fact,  these  changes  are  considered  as  important  precritical
phenomena  of  QCD  phase  transition  in  RHIC  QGP  experiments,  and
corresponding lattice QCD calculations have been performed at quenched
level~\cite{taro,umeda}.
In  this  paper, we study the thermal $0^{++}$ glueball,  whose mass 
reduction seems natural  in terms of the large difference between 
the glueball mass $m_{\rm G} \simeq$ 1.5GeV and 
$T_c \simeq$ 280MeV in quenched QCD~\cite{ichie,ishii3}.

To study the  pole mass of a  hadron in lattice QCD, one  has first to
construct  a  temporal correlator  as  $G(\tau)  = \langle  \phi(\tau)
\phi(0) \rangle$, and then resorts to its spectral representation as
\begin{equation}
  G(\tau)
=
  \int_0^{\infty} d\omega
  K(\tau,\omega)
  A(\omega),
\label{spectral.rep}
\end{equation}
where      $K(\tau,\omega)      \equiv      \frac{\cosh(\omega(\beta/2
  -\tau))}{\sinh(\beta\omega/2)}$, $\beta \equiv 1/T$, and $A(\omega)$
is the spectral function with  its spatial momentum projected to zero,
i.e., $A(\omega) \equiv  A(\omega,\vec{p}=\vec 0)$. Each peak position
of $A(\omega)$  provides us  with a  pole mass of  a hadron  at $T>0$,
which can be observed in high-energy experiments.

The most popular way to extract the spectral function $A(\omega)$ from
the constructed correlator is to  adopt an ansatz for $A(\omega)$, and
then to perform the
fitting analysis.
In  the case  of glueballs,  we first  investigated their  pole masses
using    the   simple   narrow-peak    ansatz   with    $A(\omega)   =
C\left\{\delta(\omega -  m) -  \delta(\omega + m)\right\}$,  where $C$
and  $m$ are  the strength  and the  temperature-dependent  pole mass,
respectively  \cite{ishii}. 

We then  proceeded to an  advanced analysis adopting  the Breit-Wigner
ansatz, a more general and  sophisticated ansatz, for the shape of the
spectral function as \cite{ishii2}
%
%
\begin{eqnarray}
A(\omega)    =    C\left\{\delta_{\Gamma}(\omega    -   \omega_0)    -
\delta_{\Gamma}(\omega + \omega_0)\right\}
\end{eqnarray}
expressed with the peak function of 
\begin{eqnarray}
\delta_{\Gamma}(\omega)\equiv
\frac1{\pi}\frac{\Gamma}{\omega^2  +  \Gamma^2}.
\end{eqnarray}
Here, $C$, $\omega_0$ and $\Gamma$ are the strength, 
the  peak center corresponding to the pole mass, and the thermal width
of  the glueball  peak, respectively.
(The thermal width should not be confused with the decay  width. It is
important to keep in mind that, at $T>0$, the thermal width is generated
even for  stable particles through the interaction  with the thermally-excited particles.)

The Breit-Wigner  analysis  indicates  that  the main thermal  effect for the
$0^{++}$ glueball appears as  the  thermal width broadening~\cite{ishii2} of about  300 MeV 
with a reduction in the peak center of about 100 MeV.
%
This Breit-Wigner ansatz for the lowest-lying peak in the spectral function  
seems natural and solid at relatively low temperature. 
However, near and above the critical temperature, 
more complicated structure may appear  in the spectral function. 

In this point, Maximum Entropy Method (MEM) is quite attractive, because  it provides us with
a numerical procedure to  reconstruct the spectral function $A(\omega)$ directly from
lattice QCD  Monte Carlo calculations~\cite{jarrell,asakawa}.
In this paper, we present  our preliminary  results on  the MEM-reconstructed
$0^{++}$ glueball spectrum at $T>0$.

We first consider the smearing method. Use of the extended operator is
one of the  most useful techniques to enhance  the low-energy spectra.
However,  it  has  a  disadvantage  that  it  may  make  the  physical
interpretation  of resulting  peak  less trivial.   It  can create  an
unphysical  bump structure even  in  the spectral  functions of  non-interacting
particles~\cite{karsch}.
For instance,  we consider the extreme case for glueballs. The  naive continuum
limit  of the  $0^{++}$ glueball  operator is  given as  $\phi \propto
G_{ij}^a  G_{ij}^a$.  After  the  smearing, it  is spatially  extended
according to  the Gaussian distribution  as
\begin{equation}
  \phi'
\propto
  \renewcommand{\arraystretch}{1.5}
  \Bs
  \begin{array}[t]{l}\displaystyle
  \int {d^3y d^3 z \over (2\pi)^{3/2} \rho^3}
  \exp\left( -{(\vec y - \vec z)^2 \over 2\rho^2} \right)
  G^a_{ij}(\vec  y)  G^a_{ij}(\vec z),
  \end{array}
\end{equation}
%
where   $\rho$   controls   the   size  of   the   spatial   
extension~\cite{ishii2}. We adopt here the Coulomb  gauge to obtain the
continuum expression.
We calculate  the corresponding two-point  functions perturbatively up
to $O(\alpha_s^0)$, and extract their spectral functions.  While
the  non-smeared  spectral function  is  given  as $A(\omega)  \propto
\omega^4$,  the  smeared  spectral  function  acquires  an  additional
Gaussian factor as
\begin{equation}
  A(\omega)
\propto
  \omega^4 \exp\left\{ -(\omega\rho)^2 / 4 \right\},
\label{smeared.spectral.function}
\end{equation}
which possesses  an unphysical bump at $\omega =  2\sqrt{2}/\rho$. The
smearing  method  suppresses  the   overlap  to  the  higher  spectral
components, and, as a consequence, this factor appears.


The  actual low-lying  glueball is  not  a perturbative  system but  a
definite bound state/resonance in quenched QCD below $T_c$. Note that,
in order to study the mass  and the width of a difinite resonance, the
problem of unphysical  bump is not serious, because  the pole position
in  the   complex  $\omega$  plane  is  unaffected   by  the  smearing
\cite{ishii2}.  In this  paper, we use the smearing  method to enhance
the lowest-lying $0^{++}$ glueball  peak, which is negligibly small in
the non-smeared spectral function.

We use the SU(3) anisotropic lattice plaquette action~\cite{klassen} as
 \begin{equation}
   S_{\rm G}
 =
   \renewcommand{\arraystretch}{1.5}
   \Bs
   \begin{array}[t]{l} \displaystyle
   {\beta_{\rm lat}\over N_c}\frac1{\gamma_{\rm  G}}
   \sum_{s,i<j\le  3}\mbox{Re}\mbox{Tr}
   \left\{  1  -  P_{ij}(s)  \right\}
   \\\displaystyle
   +
   {\beta_{\rm lat}\over N_c}
   \gamma_{\rm G}
   \sum_{s, i,j   \le  3}\mbox{Re}\mbox{Tr}
   \left\{ 1  - P_{i4}(s)\right\},
   \end{array}
 \end{equation}
where $P_{\mu\nu}(s) \in  \mbox{SU}(3)$ denotes the plaquette operator
in the $\mu$-$\nu$-plane.
The lattice parameter and the  bare anisotropic parameter are fixed as
$\beta_{\rm  lat}  \equiv  2N_c/g^2  =  6.25$ and  $\gamma_{\rm  G}  =
3.2552$, respectively, so as  to reproduce the renormalized anisotropy
as $a_s/a_t = 4$.  The scale  unit is introduced from the on-axis data
of  the   static  inter-quark   potential  with  the   string  tension
$\sqrt{\sigma} = 440$ MeV. The resulting lattice spacings are given as
$a_t^{-1} = 9.365(66)$ GeV  and $a_s^{-1}=2.341(16)$ GeV. The critical
temperature   is  estimated  as   $T_c  \simeq   280$  MeV   from  the
susceptibility of  the Polyakov  loop. We generate  5,500--9,900 gauge
configurations   to   construct   the  glueball   correlators,   where
statistical  data are  divided into  bins of  the size  100  to reduce
possible auto-correlations near the critical temperature.  We adopt an
appropriate smearing to enhance the low-energy spectrum.

To reconstruct the spectral function  $A(\omega)$, we apply MEM to the
appropriately  smeared  glueball  correlator normalized  according  to
$G(\tau=0)  =  1$.   As  the  practical  numerical  procedure,  we  in
principle follow \Ref{asakawa}.  We adopt the following Shannon-Jaynes
entropy as
\begin{equation}
  S
\equiv
  \int_0^{\infty}
  \left[
    A(\omega)
    - m(\omega)
    - A(\omega)\log\left({A(\omega)\over m(\omega)}\right)
  \right],
\end{equation}
where $m(\omega)$  is a real and  positive function referred  to as the
default  model  function.   $m(\omega)$   is  required  to  mimic  the
asymptotic  behavior  of  $A(\omega)$  as  $\omega  \to  \infty$.   As
$m(\omega)$,   we  adopt  \Eq{smeared.spectral.function},   i.e.,  the
perturbative spectral function up to $O(\alpha_s^0)$ as
\begin{equation}
  m(\omega)
=
  N \omega^4 \exp\left\{ -(\omega\rho)^2/4 \right\},
\end{equation}
where  the normalization  factor  $N$  is determined  so  as to  mimic
$G(\tau=0)=1$, i.e.,
\begin{equation}
  1 = \int_0^{\infty} d\omega K(\tau=0,\omega) m(\omega).
\end{equation}

\begin{figure}
\includegraphics[width=0.345\textwidth,angle=-90]{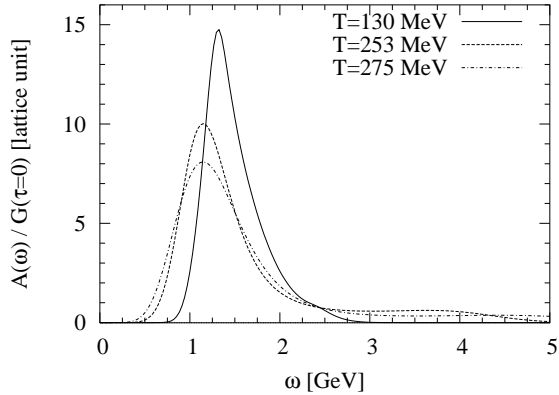}
\caption{Reconstructed spectral functions of the lowest $0^{++}$ glueball from the temporal correlators 
at $T=130, 253, 275$ MeV.}
\label{figure}
\end{figure}
In  \Fig{figure}, we  show  the reconstructed  spectral functions  
of the lowest $0^{++}$ glueball from the temporal correlations at 
$T=130, 253, 275$ MeV.
Since the error bar estimated by following \Ref{asakawa} appears to be
unreasonably  small,  we  do  not  put it  in  \Fig{figure}  to  avoid
unnecessary confusion.  For a  reasonable estimate, it seems necessary
to use the jackknife error estimate~\cite{sasaki}.
In \Fig{figure},  we see  the tendency that  the peak  becomes broader
with increasing temperature.

To  summarize,  we  have  applied  Maximum  Entropy  Method  (MEM)  to
appropriately  smeared  glueball   correlators  constructed  with  the
anisotropic SU(3)  lattice QCD at finite temperature  below $T_c$, and
have presented our preliminary results.  We have seen the tendency that the
peak becomes  broader with increasing temperature,  which supports the
thermal width broadening~\cite{ishii2} of the glueball near the critical temperature.
It is interesting to apply MEM to the glueball correlator above $T_c$,
where it becomes less trivial to find a proper ansatz for the spectral
function.
Note that  there may survive some non-perturbative  effects even above
the critical temperature.
In fact,  it was  suggested, in $\sigma$-$\pi$  sector, that  a strong
correlation   may  still  survive   above  the   critical  temperature~\cite{detar}.
To  investigate  the  correlation  above  $T_c$ with MEM, it would be necessary to analyze 
the non-smeared correlators.

\begin{center}{\bf Acknowledgment}\end{center}
N.I. thanks Dr. S.~Sasaki for useful information.

\end{document}